
\input phyzzx
\hoffset=0.375in
\overfullrule=0pt

\twelvepoint
\font\bigfont=cmr17
\centerline{   }
\centerline{\bigfont Lunar Occultation of MACHOs}
\smallskip
\centerline{\bf Cheongho Han}
\centerline{\bf Vijay K. Narayanan}
\centerline{\bf Andrew Gould$^{1}$}\foot{$^{1}$ Alfred P. Sloan Foundation
Fellow}
\smallskip
\centerline{Dept.\ of Astronomy, The Ohio State University, Columbus, OH 43210}
\smallskip
\centerline{e-mail cheongho@payne.mps.ohio-state.edu}
\centerline{e-mail vijay@payne.mps.ohio-state.edu}
\centerline{e-mail gould@payne.mps.ohio-state.edu}
\bigskip
\centerline{\bf Abstract}

\doublespace

Lunar occultation can be used to measure the proper motions of some of
the long time scale microlensing events, $t_{e} \gsim 70$ days, now being
detected
toward the Galactic bulge.
The long events are difficult to explain within the context of standard
models of the mass distribution and dynamics of the Galaxy.
Han \& Gould (1995b) have suggested that they may be due to a kinematically
cold population near the Sun.
To resolve the mass, distance, and velocity of individual events and so
to determine their nature, one must measure parallaxes and proper motions.
For long events, parallaxes can be often obtained from ground-based
measurements, but proper motions can only rarely be determined using
conventional methods.
Lunar occultations are therefore key to the understanding of the long events.
We carry out realistic simulations to estimate the uncertainty of these
measurements and show that proper motions could be measured for
about one long event per year.


\endpage

\chapter{Introduction}
\doublespace

The MACHO (Alcock et al.\ 1995; Bennett et al.\ 1995) and OGLE
(Udalski et al.\ 1994) collaborations have reported 54 candidate lensing
events toward the Galactic bulge caused by Massive Compact Objects (MACHOs).
The current estimate of the optical depth obtained by both groups is
significantly
higher than the theoretical estimate (Griest et al.\ 1991).
Various solutions have been proposed to explain the observed optical depth
excess.
One solution assumes a bar-shaped Galactic bulge
(Kiraga \& Pacy\'nski 1994; Zhao, Spergel, \& Rich 1994).
However, the excess optical depth is not completely solved even with the
adoption of a triaxial bulge (Han \& Gould 1995a).

Recently, Han \& Gould (1995b) argued that the long events, those with time
scales
of $t_{e} \gsim 70\ {\rm days}$, cannot be explained
within the context of the standard models of the mass distribution and stellar
dynamics of the Galaxy, leaving the origin of these events as a puzzle.
Although only three of the reported 54 events observed belong to this group,
these long events contribute a significant fraction to the optical depth,
provided that they are truly caused by MACHOs.
Hence, their potential importance far exceeds their number.
Han \& Gould (1995b) proposed that long events might be caused by a dynamically
cold population of dark objects such as black holes or neutron stars.
Being dynamically cold, such objects would be  located very close to the
Galactic plane, and thus only objects near the Sun could contribute to
the events detected toward Baade's window located at $b=-3^{\circ}\hskip-2pt
.9$.
The objects would then have low transverse speeds and so long $t_{e}$.

The time scale, $t_{e} $, is the only observable from current observations,
but from $t_{e}$ alone it is difficult to constrain the physical
parameters of the long $t_{e}$ events.
The time scale is related to the physical parameters by
$$
t_{e} = {r_{e} \over v};\ \
r_{e}^{2} = {4GM \over c^{2} } {D_{\rm ol}D_{\rm ls} \over D_{\rm os} }.
\eqno(1.1)
$$
Since there are three parameters, and only one measured quantity, the
individual events are highly degenerate.
Here $r_{e}$ is the Einstein ring radius, $M$ is the MACHO mass, $v$ is its
transverse
speed relative to the observer-source line of sight, and $D_{\rm ol}$, $D_{\rm
os}$,
and $D_{\rm ls}$ are the distances between the observer, lens, and source.

There have been two general approaches to break the degeneracy of the physical
parameters; parallax and proper motion measurements.
If the parallax is measured one can obtain the projected speed $\tilde{v}$:
$$
\tilde{v} = {D_{\rm ol} \over D_{\rm ls} } v.
\eqno(1.2)
$$
The proper motion, $\mu$, is determined by measuring the angular size of the
Einstein ring, $\theta_{e}$:
$$
\mu = {\theta_{e} \over t_{e} };\ \
\theta_{e} \equiv {r_{e} \over D_{\rm ol} }.
\eqno(1.3)
$$
If both the proper motion and parallax of a MACHO are measured, one can
obtain the distance to the MACHO by
$$
D_{\rm ol} = D_{\rm os}\left( {\mu \over \tilde{v}} D_{\rm os}
+1 \right)^{-1},
\eqno(1.4)
$$
since the distance to the source stars $D_{\rm os}=8\ {\rm kpc}$ is known.
Similarily $M=(c^{2}/4G)t_{e}^{2}\tilde{v} \mu$ and
$v^{-1} = \tilde{v}^{-1} + (\mu D_{\rm os})^{-1}$.
Therefore, the physical parameters of MACHOs would be uniquely determined
from the combined information of parallax and proper motion.

For typical short events seen toward the bulge, parallax can be measured
only from space (Gould 1994, 1995), and proper motions can be measured
for only a few percent of events where the MACHO transits the face of the star
(Gould 1994; Nemiroff \& Wickramasinghe 1994; Witt 1995).
However, parallax measurements for long-time-scale events become feasible
from the ground while proper motion measurement with traditional techniques
becomes essentially impossible.
Ground based parallax measurements are possible because the Earth moves through
a
substantial fraction of the Einstein ring during the event.
Indeed Bennett et al. (1995) measured a parallax for one of the three
long events using only the routine monitoring data (i.e., no
followup photometry).
On the other hand, proper motions from transits are rare because
$\theta_{e} \gsim 1\ {\rm mas}$ while the angular source size
$\theta_{\star} \lsim 10\ {\rm \mu as}$, even for giants.
However, large $\theta_{e}$ opens a possibility of resolving the two images of
a
lensed star and measuring their angular separation, $\phi_{\rm sep}$.
Once $\phi_{\rm sep}$ is measured, one can uniquely determine $\mu$ (see \S 2).
Due to the large $\theta_{e}$ of long events, it may one day be possible
to measure $\phi_{\rm sep}$ with high-resolution interferometry.
Unfortunately, current interferometry has not achieved high enough resolution
to resolve the expected very small separations of images; $\phi_{\rm sep} \sim
3\ {\rm mas}$ for long events (see \S 3.2).
Instead of interferometry, one can measure $\phi_{\rm sep}$ by using
lunar occultation.
We show in \S 5.2 one can only measure the angular
separation along a certain direction.
However, this uncertainty can often be resolved with information about the
lens geometry provided by parallax measurements.
Even when the information from parallax measurement is inadequate, two
lunar occultation measurements can resolve the degeneracy.
Hence, lunar occultation gives enough information to determine the proper
motion
and so $M$, $v$, and $D_{\rm ol}$.

In this paper we analyze the possibility of measuring the angular separation of
images of a gravitationally-lensed star using the lunar occultation method and
determine the uncertainty of the measurement by carrying out realistic
simulations.
We find that if the observation is carried out for bright, moderate-highly
magnified Galactic bulge giant stars, one can measure $\phi_{\rm sep}$ with an
uncertainty ${\mit\Delta} \phi_{\rm sep} \lsim 1\ {\rm mas}$.
The observational strategy and the method to find the intrinsic angular
separation from the measured separation normal to the Moon's surface are
discussed in \S 5.

\chapter{Proper Motion From Angular Separation Measurement}

The angular Einstein ring radius and therefore the proper motion
of the long events can be measured if the angular separation of the two
images of a lensed star is measured.
The locations of the images of the star are the solution to the quadratic
lens equation given by
$$
\theta_{\rm I} - \theta_{\rm S}\theta_{\rm I} - \theta_{e}^{2} = 0,
\eqno(2.1)
$$
where $\theta_{\rm I}$ and $\theta_{\rm S}$ are the image and source angle
measured from the lens location.
Then the angular separation of the two images is given by
$$
\phi_{\rm sep}  =
\left\vert \theta_{I+} - \theta_{I-} \right\vert =
(x^{2} + 4)^{1/2} \theta_{e},
\eqno(2.2)
$$
where the dimensionless parameter $x$ is the separation between the source and
lens
measured in units of $\theta_{e}$: $x \equiv \theta_{\rm S} / \theta_{e}$.
The value of $x$ is uniquely determined from the light curve
because the magnification of the lensed star is related to $x$ by
$ A(x) = (x^{2} + 2) / x(x^{2} + 4)^{1/2}$.
Thus, once the angular separation between two images of the source star
is determined, one can obtain $\theta_{e}$.

\chapter{Diffraction Pattern by Lunar Occultation}

\section{Fresnel Diffraction}

A stellar image produces a Fresnel diffraction pattern
when it is occulted by the Moon.
In this case, the star is idealized as a point source and the
Moon's disk as a semi-infinite plane.
When images of a lensed star are blocked by the Moon,
the observed diffraction pattern will differ slightly from a perfect
point-source pattern because the image of the lensed star is
composed of two images with a small separation.
By carefully analyzing the occultation of a lensed star, one can use this
difference to measure the angular separation which cannot be resolved
using current telescope technology.
Fortunately, the Moon passes through the Galactic bulge
every month and this permits one to apply the lunar occultation
method to measure the separation of two images of a star lensed by a
significant fraction of MACHOs.
In addition, since very long events are suspected to be caused by MACHOs
close to the Sun, the angular separations would be large enough to
measure $\phi_{\rm sep}$ precisely by the lunar occultation method.

The diffraction pattern produced by a background source star when it is
occulted
by an opaque object (e.g., the Moon) is given by
$$
{g \over g_{0} } = {1 \over 2} \left\{ \left[ {1\over 2} - {\cal C}(z)
\right]^{2}
+ \left[ {1\over 2} - {\cal S}(z) \right]^{2} \right\},
\eqno(3.1.1)
$$
where ${\cal C}$ and ${\cal S}$ are the Fresnel cosine and sine integrals and
$g_{0}$ is the intensity without diffraction (e.g.,\ Hecht \& Zajac 1979).
The two Fresnel integrals are defined by
$$
{\cal C}(z) = \int_{0}^{z} \cos \left( {\pi \over 2} z'^{2} \right) dz',\ \ \
{\cal S}(z) = \int_{0}^{z} \sin \left( {\pi \over 2} z'^{2} \right) dz'.
\eqno(3.1.2)
$$
Here the dimensionless distance variable $z$ is defined by
$$
z = r \left[ { 2(d_{\rm OM} + d_{\rm M}) \over \lambda d_{\rm OM} d_{\rm M}}
\right]^{1/2},
\eqno(3.1.3)
$$
where
$\lambda$ is the wavelength of the observation,
$d_{\rm M}$ and $d_{\rm OM} = D_{\rm os} - d_{\rm M}$ are the distance to the
Moon from the Earth and from the source star,
and $r=d_{\rm M}\theta$.
Since $d_{\rm OM} \gg d_{\rm M}$, one can approximate eq.\ (3.1.3) by
$ z = r (2/ \lambda d_{\rm M} )^{1/2}$.
When the image of a star is directly on the edge of the Moon's surface,
$v=r=0$,
${\cal C}(0) = {\cal S}(0) = 0$ and $g / g_{0} = 0.25$.

\section{Simulation}

In our simulation the observations are assumed to be carried out as follows.
During an occultation event, photometry is carried out continuously
(i.e., with time resolution $\ll 1\ {\rm mas}$) and read instantaneously.
This kind of high speed photometry technique has already been developed and
the actual instrument (HSP) had been installed at the Hubble Space Telescope
(HST)
although the instrument was removed from HST to accommodate the installation of
corrective optics.
A large ($\gsim 4\ {\rm m}$) telescope is required for the measurement
to compensate for the relative faintness of even giant sources in the bulge.
For an $H=0$ star, a (single channel) photometer can detect
$\sim 9.4 \times 10^{9}\ \eta {\mit\Delta} \lambda\ {\rm photons}\
{\rm m}^{-2}{\rm s}^{-1}{\rm \mu m}^{-1}$.
With a band width of ${\mit\Delta}\lambda=0.3\ {\rm \mu m}$ and assuming
a detection efficiency $\eta = 0.5$, it can detect
$1.6\times 10^{7}\ {\rm photons}\ {\rm ms}^{-1}$
using a 4 m telescope with 90 \% effective surface area.
The expected number of photons from a bulge clump giant, typically
$H=13.2\ {\rm mag}$ (Tiede, Frogel, \& Terndrup 1995), will be $\sim 67$
${\rm ms}^{-1}$ assuming an extinction of $A_{H}= 0.25$ mag toward the bulge.
The sky has a brightness of $14\ {\rm mag}\ {\rm arcsec}^{-2}$ in the $H$ band
producing an expected sky flux of $71\ {\rm photons}\ {\rm ms}^{-1}$
in an aperture of 1.5 arcsec in diameter.
We assume that the occultation occurs when the dimensionless impact parameter
$x = 0.5$, and thus the magnification $A = 2.18$.
Then the individual magnifications of the image are
$$
A_{\pm} = { x^{2}_{\pm} \over x_{+}^{2} - x_{-}^{2} };\ \
x_{\pm} = { \sqrt{x^{2} + 4} \pm x \over 2},
\eqno(3.2.1)
$$
giving $A_{+}=1.59$ and $A_{-}=0.59$ for the primary and secondary
images, respectively.
The photon counts for each image are then
$N_{\nu,1} = 112\ {\rm photons}\ {\rm ms}^{-1}$
and $N_{\nu,2} = 41\ {\rm photons}\ {\rm ms}^{-1}$.
The expected number of detected photons from the individual images and from the
sky
for various observational situations are computed and listed in Table 1.
In the table the signal-to-noise ratio is computed by
$ S/N = N_{\nu} / (N_{\nu,1}+N_{\nu,2})^{1/2}$.
We assume that other sources of noise,  e.g., dark current and read-out noise,
are negligible.



The theoretically-expected fringe pattern $g/g_{0}$ in the $H$ band, centered
at
$1.65\ {\rm \mu m}$, is computed by eq.\ (3.1.1) and is shown as a function
time $t$ in Figure 1 (a).
In the figure, time is measured from the moment when the first
(closer to the Moon) image just crosses the Moon's limb and negative time
implies that the image is behind the Moon.
The angular distance is related to time scale by
$$
\theta = \omega t\cos \psi,
\eqno(3.2.2)
$$
where
$\omega$ is the orbital angular speed of the Moon and $\psi$ is the angle of
the
lunar limb relative to the direction of the Moon's motion.
The total number of photons in the signal is normalized into unity when
$t=\infty$.
Due to the phase shift of the Fresnel integrals with time, there are beat
patterns.
In the simulation, we assume the angular separation between the two images
of lensed stars is $\phi_{\rm sep} = 3\ {\rm mas}$,
equivalent to an Einstein radius $\theta_{e} \sim 1.5\ {\rm mas}$,
which would be typical for a disk lens located at $\sim 2\ {\rm kpc}$
with mass of $M=0.7\ M_{\odot}$.
Since the long events are probably caused by relatively massive MACHOs, the
expected
separation could be greater.

Up until now, our computation has been based on a monochromatic wave
observation.
However, the $H$-band filter has a finite bandpass.
For the correction of band width in our analysis, we average the flux
weighted by the filter function.
The filter function in the $H$ band is well-approximated by the tophat function
$\omega_{\rm filter}$:
$$
w_{\rm filter} =
\cases{\omega_{0}, &  $ 1.5\ {\rm \mu m} \leq \lambda \leq 1.8\ {\rm \mu m}$\cr
0, & otherwise,\cr}
\eqno(3.2.3)
$$
where $\omega_{0} = (0.3\ {\rm \mu m})^{-1}$.
The resultant fringe pattern $h(\theta)$, after being averaged by the filter
function, is shown in Figure 1 (b).
The small-scale fluctuation in flux is smeared out, and thus the beat patterns,
which might have been useful for the measurement of $\phi_{\rm sep}$,
disappear.

We make the simulation more realistic by including the beam pattern of the
telescope.
At any time, one edge of the telescope mirror will see an image displaced by
a small angle compared to the image seen at the center of the mirror.
The resulting image is then the combination of all images seen
by different parts of the mirror.
The beam pattern decreases with increasing angular separation $\theta_{\rm p}$
from
the center of the mirror because the relative area of the mirror in a strip
located at $\theta_{\rm p}$ decreases as $\theta_{\rm p}$ increases.
For a circular mirror
$f_{\rm beam}=\left[ 1-(\theta_{\rm p} /\theta_{\rm tel})^{2}\right]^{1/2}$,
where $\theta_{\rm tel} =  a/2d_{\rm M} = 1\ {\rm mas}$ for an
assumed aperture $a=4\ {\rm m}$.
Then the expected intensity $h(\theta)$ is obtained by convolving
the intensity $h(\theta)$ with the beam pattern
$f_{\rm beam}(\theta_{\rm p} )$:
$$
I(\theta) = \int f_{\rm beam}(\theta_{\rm p} - \theta)
h(\theta_{\rm p} ) d\theta_{\rm p} .
\eqno(3.2.4)
$$
The resulting fringe pattern after the beam pattern convolution and being
averaged
by filter function is shown in Figure 1 (c).

\chapter{Uncertainty of Measurement and Observational Strategy}

The determination of the angular separation of the two images of a lensed star
is made by fitting the observed fringe pattern to a set of light curves with
different $\phi_{\rm sep}$.
This fitting process suffers from another uncertainty in addition to the one
from
the angular separation of the images.
This uncertainty comes from the ambiguity of a reference point, which is the
time of first image occultation, (i.e., $\theta = 0$).
A misalignment of the reference point would result in misinterpretation of
$\phi_{\rm sep}$.
Therefore, we include the misalignment quantified by the shift in the reference
point, ${\mit\Delta} t_{0}$, as a free parameter in our analysis.

The contours of equal uncertainties measured by $\chi^{2}$ in the
parameter space of ${\mit\Delta} t_{0}$ and $\phi_{\rm sep}$ are shown in
Figure 2 (a) for the first lensing event described in Table 1.
The contour levels are drawn at $1\ \sigma$, $2\ \sigma$, and $3\ \sigma$
levels
from the point of minimum $\chi^{2}$ marked by ``x''.
In the contour map there appear two minima which are located symmetrically
about $\phi_{\rm sep} = 0$.
These two minima appear because of the ambiguity of order of occultation:
the fringe pattern when the primary image approaches the Moon first looks
similar to the pattern when the other image approaches first.
If the two images have exactly the same intensity, there will be  no
difference in the resultant fringe patterns.
The uncertainty is, ${\mit\Delta} \phi_{\rm sep} \sim 0.55\ {\rm mas}$
at the $1\ \sigma$ level.
This would be a significant detection although not as precise as one
might like.

However, the uncertainty decreases significantly with increasing $S/N$ ratio.
There are two ways to increase the $S/N$ ratio; observing bright stars or
highly-magnified events.
For illustration, we compute the uncertainties for events under exactly the
same
conditions as the previous case except for a higher magnification $A=3.0$
[Fig. 2 (b)] and a 0.5 mag brighter source star [Fig. 2 (c)],
which are the second and third cases in Table 1, respectively.
For both cases the uncertainty is
${\mit\Delta} \phi_{\rm sep} \sim 0.35\ {\rm mas}$ at the 1 $\sigma$ level.

\chapter{Practical Considerations}

\section{Lunar Topography}

For an actual observation, one is required to consider some miscellaneous
factors that make $\phi_{\rm sep}$ deviate somewhat from what we have
assumed in the computation.
The first deviation arises because the Moon's limb is not a perfect
straight line but has a small curvature.
If the angular separation between images is big enough (e.g., binary stars),
the observed pattern would slightly deviate from the Fresnel diffraction
pattern,
which assumes an infinite straight-line surface.
However, for the scale of $\lsim 5.5\ {\rm m}$, which is equivalent
to $\phi_{\rm sep} \lsim 3\ {\rm mas}$  projected on the Moon's
surface, the Moon's limb is a nearly perfect straight line.
Indeed, typical errors arising from lunar surface curvature are
negligible even for the measurement of a few tens of arcsec
(e.g., Richichi et al.\ 1994).
The other type of deviation which {\it does} affect on the measurement
of $\phi_{\rm sep}$ is due to topographical features on the lunar surface.
If the images are occulted at different parts of a geological structure on
the Moon, e.g., moutains and valleys,
the angle $\psi$ in eq.\ (3.2.2) deviates from what one assumes based on
a smooth circular lunar disk.
Assuming that the deviation due to large scale strucures of order km
or larger can be corrected using a detailed lunar map, one still has a problem
due to smaller-scale structures such as rocks and cliffs.
Fortunately, lunar craters and other topographical features are
generally not very rugged, and even mountains are very smooth
(Abell, Morrison, \& Wolff\ 1993).

\section{Orientation of Images}

We have assumed up until now that the orientation of the two images is
perpendicular
to the approaching limb of the Moon.
However, the orientation in general is random and thus what one measures is
the component of $\phi_{\rm sep}$ normal to the Moon's limb not the
intrinsic separation of interest.
If one knows exactly how the source star crosses the Einstein ring,
it is simple geometry to deduce $\phi_{\rm sep}$ from the measured angular
separation $\phi_{i}$, where the subscript ``$i$'' is explained below.
However, resolving the orientations of the images is not a trivial problem
and requires additional information.

Information about the  lens geometry can be obtained from (ground-based)
measurement of the parallax.
At a minimum, parallax measurements determine the component of the projected
velocity parallel to the ecliptic and the magnitude of the component
normal to the ecliptic.
In general, this still leaves a four-fold degeneracy: two-fold for the sign of
the
normal component and two-fold for the sense of source motion (clockwise or
counterclockwise) relative to observer-lens line of sight.
In some cases, especially, if an event is observed away from the ecliptic and
has
a long time scale, this four-fold degeneracy is completely broken by the
parallax measurement itself.
Indeed, this is the case for the long-event parallax measured by Bennett et
al.\ (1995).
In these cases the geometry is unambiguous and $\theta_{e}$ can be determined
directly from $\phi_{\rm sep}$.
However, it is difficult to break this degeneracy when the event is close to
the
ecliptic.
Since the Moon never gets farther than $\sim 5^{\circ}$ from the ecliptic, it
is
also important to consider the degenerate case.

The four-fold degeneracy can be broken by carrying out occultation observations
twice so that the approaching angle, $\theta_{\rm A}$, of the Moon's limb
is different for each observation.
This can be done either by observing the occultation at two different locations
on the Earth or by observing an occultation from the same location
a month later.
The latter case is possible for long events since the declination of Moon's
orbit
typically changes by $\sim 1/3$ of its diameter per month at fixed right
ascension.
The four-fold degeneracy of lens geometries is illustrated in Figure 4:
transverse velocities pointing below and above the ecilptic are noted by
``case I'' and ``case II''.
Subnotations, ``(a)'' and ``(b)'', describe the clockwise and counterclockwise
source crossings.
For illustration, we assume that the Moon's limb is approaching the images with
$\theta_{\rm A}=45^{\circ}$ (``/'' sense) and
$\theta_{\rm A}=-45^{\circ}$ (``$\setminus$'' sense).
For each observation shown in the figure
the shaded surface indicates the orientation of the surface of the Moon
and the shaded arrow indicates the direction of motion.
The angular separation that is measured when $\theta_{\rm A}=45^{\circ}$
and $\theta_{\rm A}=-45^{\circ}$ is marked by $\phi_{1}$ and $\phi_{2}$.
For each case the ratios between measured angular separations determined at
two different locations (or times), $\phi_{1}/ \phi_{2}$, are
different from one another.
Therefore, one can obtain the full lens geometry by ruling out the other
three cases from a comparison of the observed ratio $\phi_{1}/ \phi_{2}$
with the expected ratios.

\section{Event Rate}

One can measure proper motions using lunar occultations for about one long
time-scale
lensing event per year.
The occulation can be observed in a strip of sky along the path of the Moon
with a width larger than the angular size of the Moon by observing the
occultation at higher (or lower) Earth latitudes, $\phi$.
The effective occulting cross-section is $\sim 1^{\circ}\hskip-2pt .5$
assuming the observations could be arranged at places within
$\phi = \pm 30^{\circ}$.
Then the area of sky toward the bulge, with width $\pm 5^{\circ}$ around
the Galactic plane, swept by the effective cross-section is
$15\ {\rm deg}^{2}$.
If the observation could be carried out at more extreme Earth latitudes,
the event rate could be increased.
The MACHO group has detected 13 giant events out of a total of 44 events during
a bulge season by covering $12\ {\rm deg}^{2}$ of sky.
Three of 44 are long events.
The expected occultation event rate is then $3\times (13/44) \times
(15/12) \sim 1\ {\rm event/yr}$.
This is substantially higher than the number of proper motions of long lensing
events that can be obtained using other methods.
Another advantage of the occultation method is that for the long events the
measurement could be repeated for better determination of proper motion.
Special attention should be paid to observing those fields which the Moon
will occult sometime during a given bulge season.
For some regions close to the plane, optical observations are impossible due
to extinction.
However, events can still be detected using $H$ or $K$ band observations.
Indeed, the long events may well be concentrated near the plane (see
Fig.\ 1 from Bennett et al.\ 1995).
Because the long events last many months, it would be sufficient to make
these infrared observations once per week compared to once per day
for standard optical observations.

{\bf Acknowledgement}: We would like to thank M. Everett and G. Newsom for
making very helpful comments.

\endpage
\ref{Abell, G. O., Morrison, D., Wolff, D. C.\ 1993, Exploration of the
Universe
(Asunder College Publishing, Philadelphia), 222}
\ref{Alcock, C. et al.\ 1995, ApJ, 445, 133}
\ref{Bennett, D. P. et al.\ 1995, Proceedings of the Fifth Annual Maryland
Meeting on
Astrophysics: Dark Matter, S. S. Holt, C. L. Bennett, eds., in press}
\ref{Gould, A. 1992, ApJ, 392,442}
\ref{------------- 1994, ApJ, 421, L71}
\ref{------------- 1995, ApJ, 447, 000}
\ref{Gould, A., Miralda-Escud\'e, J., \& Bahcall, J. N.\ 1994, ApJ, 423, L105
}
\ref{Griest, K. et al. 1991, ApJ, 387, 181}
\ref{Han, C., \& Gould, A.\ 1995a, ApJ, 447, 000}
\ref{-----------------------------  1995b, ApJ, submitted}
\ref{Hecht, E. \& Zajac, A.\ 1979, Optics (Addison-Wesley Publishing
Company, Reading), 385}
\ref{Kiraga, M., \& Paczy\'nski, B.\ 1994, ApJ, 430, L101}
\ref{Nemiroff, R. J. \& Wickramasinghe, W. A. D. T.\  1994, ApJ, 424, L21}
\ref{Richichi, A., Calamai, G., \& Leinert, C. 1994, A\&A, 286, 829}
\ref{Tiede, G. P., Frogel, J. A., \& Terndrup, D. M.\ 1995, AJ, submitted}
\ref{Udalski, A., Szyma\'nski, M., Stanek, K. Z., Kalu\.zny, J., Kubiak, M.,
Mateo, M., Krzemi\'nski, B., \& Venkat, R.\ 1994, Acta Astron., 44, 165}
\ref{Witt, H.\ 1995, ApJ, in press}
\ref{Zhao, H., Spergel, D. N., \& Rich, R. M.\ 1995, ApJ, 440, L13}

\refout
\endpage
\endpage
\bye